\documentstyle[11pt]{article}
\def\init{\setcounter{equation}{0}}

\newtheorem{theorem}{Theorem}[section]
\newtheorem{proposition}[theorem]{Proposition}

\newtheorem{corollary}[theorem]{Corollary}

\def\rr{{\bf R}}

\def\E{{\bf E}}

\def\calf{{\cal F}}

\def\calh{{\cal H}}

\def\calx{{\cal X}}
\def\calth{{\tilde{\calh}}}

\def\tg{\tilde g}

\def\tH{\tilde H}

\def\hp{\tH_{\psi}}

\def\ie{{\it i.e., }}
\def\eg{{\it e.g., }}

\def\qed{$\Box$}

\def\half{\frac{1}{2}}

\def\proof{\noindent{\it  Proof. }}
\def\p{\partial}

\def\drs{\frac{\p}{\p r_*}}
\def\dds{\frac{d}{ds}}
\def\Drs{D_{r_*}}
\def\rs{{r_*}}
\def\bl{\Big(}
\def\br{\Big)}

\def\intt{\int_{-\infty}^\infty}
\def\eps{\epsilon}

\def\supp{\hbox{supp}\,}
\def\jap#1{\langle{ #1} \rangle}

\newcommand{\beq}{\begin{equation}}
\newcommand{\eeq}{\end{equation}}

\begin{document}

\title{Global existence and scattering for the nonlinear
Schr\"odinger equation on Schwarzschild manifolds}
\author{I. \L aba\\ Department of Mathematics\\ Princeton
University\\Princeton, NJ 08544\\ {\it laba@math.princeton.edu}
\and A. Soffer\\ Department of Mathematics\\ Rutgers University\\
Piscataway, NJ 08854\\ {\it soffer@math.rutgers.edu}}
\date{September 20, 1999 (revised)}

\maketitle

\begin{abstract}
We consider the nonlinear Schr\"odinger equation with a pure power
repulsive nonlinearity on Schwarzschild manifolds. Equations of this
type arise when a nonlinear wave equation on a Schwarzschild
manifold is written in Hamiltonian form, cf. \cite{BaNi}, \cite{Ni}.
For radial solutions with sufficiently localized initial data, we
obtain global existence, $L^p$ estimates, and the existence
and asymptotic completeness of the wave operators. Our approach is
based on a dilation identity and global space-time estimates.



\end{abstract}


\section{Introduction}
\label{intro}
\init


A {\em Schwarzschild manifold} is the space $\rr\times\rr^+\times S^2$
equipped with the Schwarzschild metric:
\[
g=g_{\mu\nu}dx^\mu dx^\nu,
\]
which may be written in polar coordinates as:
\beq
g =\bl 1-\frac{2M}{r}\br dt^2-\frac{dr^2}{1-\frac{2M}{r}}
-r^2\Delta_{S^2};
\label{ds.e1}
\eeq
$\Delta_{S^2}=d\theta^2+\sin^2\theta d\phi^2$
is the Laplace-Beltrami operator on a 2-dimensional sphere.
The parameter $M>0$ is interpreted as the mass of the black hole.
We restrict our attention to the {\it external} Schwarzschild solution 
($r>2M$).

Scattering theory for the wave and Klein-Gordon equations on
Schwarzschild manifolds was first studied by Dimock \cite{Di} and
Dimock and Kay \cite{DiKa}. Following these authors, we rewrite
(\ref{ds.e1}) as:
\beq
g =\bl 1-\frac{2M}{r}\br (dt^2-dr_*^2)-r^2\Delta_{S^2},
\label{ds.e2}
\eeq
where $\rs$ is the Regge-Wheeler tortoise radial coordinate:
\beq
\rs=r+2M\log(r-2M)
\label{e.rs}
\eeq
(hence $\frac{dr}{d\rs}=1-\frac{2M}{r}$).
The wave equation on the Schwarzschild manifold is:
\beq
\Box_g u=0,
\label{box.e1}
\eeq
where:
\beq
\Box_g= |\det g|^{-1/2}\frac{\p}{\p x^\mu}
\bl |\det g|^{1/2}\,g^{\mu\nu}\frac{\p}{\p x^\nu}\br
\label{box.e2a}
\eeq
is the d'Alembertian associated to $g$. Using (\ref{ds.e2}), we may
write (\ref{box.e2a}) as:
\beq
\Box_g=\bl 1-\frac{2M}{r}\br^{-1}
\bl \frac{\p^2}{\p t^2}-r^{-2}\drs r^2\drs \br
-r^{-2}\Delta_{S^2}.
\label{box.e2}
\eeq
(\ref{box.e1})--(\ref{box.e2}) is equivalent to:
\beq
\frac{\p^2}{\p t^2}u+Hu=0,
\label{box.e3}
\eeq
where:
\beq
H=-r^{-2}\drs r^2\drs -r^{-2}\bl 1-\frac{2M}{r}\br \Delta_{S^2}.
\label{box.e6}
\eeq
In \cite{Di}, it was proved that the wave operators for (\ref{box.e3})
exist and are complete. For the Klein-Gordon equation $\Box_g\,u+m^2u=0$
the existence of wave operators was proved in \cite{DiKa}, and
asymptotic completeness -- in \cite{Ba1}. The results of \cite{Di}
on scattering were recovered in \cite{DHS}, where the authors 
actually considered a more general class of noncompact manifolds;
the proof in \cite{DHS} relied on a Mourre estimate obtained 
for such manifolds in \cite{FH}. (See also \cite{SZ}, where certain
techniques of geometric scattering theory were applied to the
De Sitter model.)

An important open problem is to develop a scattering theory for
equations such as (\ref{box.e3}), but with a nonlinear perturbation
added. Partial results in that direction were obtained in \cite{BaNi}, 
\cite{Ni}; the class of metrics considered there is in fact more general
than (\ref{ds.e1}) and includes other black hole models. In particular,
Nicolas \cite{Ni} studied a nonlinear Klein-Gordon equation of the form:
\[
\Box_g\,u+m^2u+\lambda F(r)|u|^2u=0,
\]
where $F(r)$ is an explicitly given factor vanishing as $r\to 2M$ and
as $r\to\infty$. He obtained the global existence of solutions to the
above equation and an outgoing radiation condition for these solutions.
We also remark that the Cauchy problem for the Yang-Mills equations on
Schwarzschild manifolds was studied in \cite{Shu}.

In the present paper, we take a different route and study the
scattering theory for the nonlinear Schr\"odinger equation:
\beq
i\frac{\p u}{\p t}=Hu+\lambda |u|^{p-1}u,\ \lambda>0;
\label{ham.e1}
\eeq
we will, moreover, restrict our attention to radially symmetric 
solutions, \ie assume that:
\beq
\Delta_{S^2}u\equiv 0.
\label{e.ssym}
\eeq
The payoff for these simplifications is that we will be able to present
a relatively elementary (modulo the estimates of \cite{W}) proof of
existence and completeness of the wave operators. As one might guess by
considering the geometry of the manifold, (\ref{ham.e1}) has two
scattering channels: part of the outgoing wave escapes to the (spatial)
infinity and becomes asymptotically free, while another part approaches
the black hole horizon and therefore displays a different asymptotic
behaviour. We prove that (in a suitable coordinate system) each part
has the asymptotics of a solution to a simpler {\em linear} equation.
Our analysis of (\ref{ham.e1}) is based on a priori estimates,
similar in spirit to the conformal and Morawetz identities;
in particular, we obtain the local decay of solutions and suitable
space-time $L^p$ estimates.  

We remark that the same proof, with only minor modifications, should
work for slightly more general nonlinearities of the form $f(|u|)u$,
where $f(s)$ is a suitable real-valued function; for the sake of
brevity, we do not attempt here to find the exact conditions on $f$
under which this can be done.

While our results do not imply anything directly about scattering for
a nonlinear {\em wave} equation (which would be more interesting than
(\ref{ham.e1}) from the point of view of physics), we believe that 
the methods presented in this article may be developed further to
yield progress in that direction.  To illustrate the connection between 
(\ref{box.e3}) and (\ref{ham.e1}), we rewrite (\ref{box.e3}) in
Hamiltonian form:
\beq
i\frac{\p}{\p t}\pmatrix{u\cr \p_t u\cr}
=\frac{1}{i}\pmatrix{0&-1\cr H&0} \pmatrix{u\cr \p_t u\cr}.
\label{box.e5}
\eeq
This was in fact the approach taken in \cite{Di}, \cite{DiKa},
\cite{Ba1}, \cite{Ni}, \cite{DHS}. By diagonalizing the matrix in
(\ref{box.e5}) one may reduce the problem to studying the unitary 
group $\exp(-it\sqrt{H})$, cf. \cite{Ba1}, \cite{DHS}. It was further
demonstrated in \cite{DHS} that certain results of this type may be
deduced from their analogues for the Schr\"odinger unitary group
$e^{-itH}$. We hope to use similar methods to make progress on the
scattering theory for a nonlinear variant of (\ref{box.e3}).

The paper is organized as follows.  Section \ref{ham} takes care
of preliminaries such as the conservation of the $L^2$ and energy
norms for the solutions of (\ref{ham.e1}).  Assuming (\ref{e.ssym}),
the problem becomes effectively one-dimensional. We simplify it
further in Section \ref{red} by applying a suitable unitary
transformation.  The point of this reduction is that the kinetic energy
part becomes simply $D^2_{\rs}$; the price we pay is that we have to 
add a ``potential" $V$. Our main estimates are proved in
Sections \ref{dil}--\ref{conf}. In Section \ref{es} we combine them
with the known $L^p$ estimates for a 1-dimensional linear Schr\"odinger
equation (see \cite{W}, where such estimates were proved and applied
to a similar nonlinear scattering problem), and obtain the time decay of
$L^p$ norms of the solutions. The global existence of solutions, and the
existence and completeness of the wave operators between the nonlinear
equation and the corresponding {\it linear} Schr\"odinger equation,
follow by a standard argument (cf. \cite{S}).

The authors are grateful to the referee for helpful remarks and
for bringing the article \cite{Ni} to their attention.
The second author acknowledges partial support by the National
Science Foundation.


\section{Preliminaries}
\label{ham}
\init


Let $u=u(r,\omega,t)$ be a solution to (\ref{ham.e1}), where $H$ is 
given by (\ref{box.e6}); we will also use the notation $u_t(\cdot)\equiv
u(\cdot, t)$. Recall that $\rs$ is defined in (\ref{e.rs}). For future
reference we note that $r$ is an increasing function of $\rs$ and:

\begin{itemize}

\item
as $\rs\to\infty$, $r\to \infty$ and $1-\frac{2M}{r}\to 1$;

\item
as $\rs\to -\infty$, $r\sim 2M+e^{-1+\rs/2M}\to 2M$ and:
\beq
1-\frac{2M}{r}\sim \frac{1}{2M}e^{-1-|\rs|/2M}
\label{ham.e2}
\eeq
vanishes exponentially in $|\rs|$.

\end{itemize}

We will assume that the initial data $u_0$ satisfies (\ref{e.ssym})
and belongs to the energy space:
\[
\calh=\{u\in L^2(\rr\times S^2;r^2d\rs\,d\omega):\ 
u(r,\omega)=u(r) \hbox{ and } \E(u)<\infty\},
\]
where the {\it energy} $\E(u)$ is defined as: 
\beq
\E(u)=\int\bl\bar u\,Hu+\frac{2\lambda}{p+1}|u|^{p+1}\br
r^2d\rs d\omega.
\label{ham.e3}
\eeq
We denote by $\|u\|_{\calh}=(\E(u))^{1/2}$ the energy norm of $u$.

Observe first that the $L^2$ norm of $u_t$ is conserved:
this follows from the fact that the operator $H$ is symmetric 
and from the form of the nonlinearity in (\ref{ham.e1}).
The second basic fact we will need is the conservation of energy.

\begin{proposition}
For any solution $u_t(r,\omega)=u(r,\omega,t)$ of (\ref{ham.e1})
(not necessarily radially symmetric), $\E(u_t)$ is independent of $t$.
In particular, if $u_0\in\calh$, then $u_t\in\calh$ and $\|u_t\|_\calh
=\|u_0\|_\calh$ for all $t$. 
\label{ham.prop1}
\end{proposition}

\proof
We compute:
\[
\begin{array}{l}
\frac{d}{dt}\int_{\rr\times S^2}
\bar u_t\cdot(H+\lambda|u_t|^{p-1})u_t\,r^2d\rs d\omega
=\int_{\rr\times S^2} \bar u_t\cdot\frac{\p}{\p t}(\lambda|u_t|^{p-1})
u_t\,r^2d\rs d\omega
\\[3mm]
=\int\frac{\p}{\p t}(\lambda|u_t|^{p-1})|u_t|^2 r^2d\rs d\omega
=\frac{p-1}{p+1}\int\frac{\p}{\p t}(\lambda|u_t|^{p+1})r^2d\rs 
d\omega.
\end{array}
\]
Hence:
\[
\frac{d}{dt}\int\bl\bar u_t\,Hu_t+\lambda|u_t|^{p+1}\br r^2d\rs 
d\omega
=\frac{\lambda(p-1)}{p+1}\,\frac{\p}{\p t}\int(\lambda|u_t|^{p+1})
r^2d\rs d\omega,
\]
which proves the proposition. \qed

\begin{corollary}
Suppose that $u_t$ solves (\ref{ham.e1}), $u_0\in\calh$. Then:
\smallskip

(i) $\int\bar u_t\,Hu_t\,r^2d\rs d\omega<C\|u_0\|_\calh^2$,

\smallskip
(ii) $\int |u_t|^{p+1}\,r^2d\rs d\omega<C\|u_0\|_\calh^2$,

\smallskip

\noindent uniformly in t.
\label{ham.cor2}
\end{corollary}

\proof 
By Proposition \ref{ham.prop1}, 
\beq
\E(u_t)=\int\bar u_t\,Hu_t\,r^2d\rs d\omega
+\frac{2\lambda}{p+1}\int |u_t|^{p+1}\, r^2d\rs d\omega
\label{ham.e5}
\eeq
is constant in $t$. Since both terms on the right-hand side of
(\ref{ham.e5}) are positive, this implies the corollary.
\qed


\section{Reduction to a one-dimensional problem}
\label{red}
\init


>From now on, we restrict our attention to the space $L^2_{radial}
(\rr\times S^2; r^2d\rs d\omega)$ of radially symmetric functions 
in $L^2(\rr\times S^2; r^2d\rs d\omega)$. We define a unitary operator
$U:L^2(\rr,d\rs)\to L^2_{radial}(\rr\times S^2; r^2d\rs d\omega)$ by:
\[
U:\psi(r)\to u(r,\omega):=r^{-1}\psi(r),
\]
and the symmetric operator $\tH$ on $L^2(\rr,d\rs)$ by:
\[
\tH=U^{-1}HU=rHr^{-1}.
\]
Using (\ref{box.e6}) and (\ref{e.ssym}), we find that:
\beq
\tH=D_\rs^2+V(\rs),
\label{red.e0}
\eeq
where $D=-i\partial$ and:
\beq
V(\rs)=\frac{2M}{r^3}\bl 1-\frac{2M}{r}\br.
\label{red.e1}
\eeq

Substituting $\psi=U^{-1}u=ru$ in (\ref{ham.e1}), we obtain that
$\psi$ satisfies::
\beq
i\frac{\p}{\p t}\psi=\hp\psi,
\label{red.e2}
\eeq
where $\hp$ is the nonlinear operator:
\beq
\hp=\tH+\lambda r^{-p+1}|\psi|^{p-1}.
\label{red.e3}
\eeq
The energy space $\calh$ is mapped by $U^{-1}$ to:
\[
\calth=\{\psi\in L^2(\rr;d\rs):\ 
\|\psi\|^2_{\calth}:=\int\bar\psi\cdot\hp\psi d\rs<\infty\}.
\]
The remark before Proposition \ref{ham.prop1}, and the unitarity of
$U$, imply the conservation of the $L^2$ norm for solutions of 
(\ref{red.e2}). Moreover, from Corollary \ref{ham.cor2} we
obtain the following.

\begin{proposition}
$U$ is a unitary operator from $\calth$ to $\calh$. Moreover, if
$\psi_t(r)=\psi(r,t)$ solves (\ref{red.e2}) and $\psi_0\in\calth$, then:

\smallskip
(i) $\int|\drs\psi_t|^2\,d\rs <C\|\psi_0\|_{\calth}^2$;

\smallskip
(ii) $\int r^{-p+1}|\psi_t|^{p+1}\,d\rs <C\|\psi_0\|_{\calth}^2$,

\smallskip
\noindent uniformly in $t$.
\label{red.prop1}
\end{proposition}

\proof
Substituting $u_t=r^{-1}\psi_t$ in Corollary \ref{ham.cor2}{\it(i)}, 
we obtain that:
\[
\begin{array}{l}
\int\overline{r^{-1}\psi_t}\cdot H(r^{-1}\psi_t)r^2d\rs d\omega
=\int\bar\psi_t\cdot rHr^{-1}\psi_t\, d\rs d\omega
\\[3mm]
=\int\bar\psi_t\cdot \tH\psi_t\, d\rs d\omega
=\int\bar\psi_t\cdot (D_\rs^2+V(\rs))\psi_t\, d\rs d\omega
\\[3mm]
=\int\bar\psi_t\cdot D_\rs^2\psi_t\, d\rs d\omega
+\int\bar\psi_t\cdot V(\rs)\psi_t\, d\rs d\omega
\end{array}
\]
is bounded by $C\|u_0\|_{\calh}^2=C\|\psi_0\|_{\calth}^2$ for all $t$. 
Moreover, since both terms in the last line are positive, we find that:
\[
\int\bar\psi_t\cdot D_\rs^2\psi_t\, d\rs d\omega<C\|\psi_0\|_{\calth}^2
\]
uniformly in $t$, which after integration by parts yields {\it(i)}.
Part {\it(ii)} follows by substituting $u_t=r^{-1}\psi_t$ in 
Corollary \ref{ham.cor2}{\it(ii)}.
\qed


\section{The dilation identity}
\label{dil}
\init


The starting point for our analysis of (\ref{red.e2}) is the
observation that both the ``potential" $V$ given by (\ref{red.e1})
and the nonlinear term in (\ref{red.e3}) are repulsive interactions.  
Hence the long-time behaviour of the solutions is largely determined
by the dispersive term $D_{\rs}^2$. In particular, we obtain the
{\em local decay} estimates (Proposition \ref{dec.thm2}); these
in turn will be needed in the proof of our results on scattering.

Throughout the rest of this paper we will denote by $\jap{\cdot,\cdot}$
the inner product in $L^2(\rr; d\rs)$: $\jap{\psi,\phi}
=\int_{-\infty}^\infty \bar\psi\phi d\rs$.

\begin{proposition}
There is an $\alpha\in\rr$ (given explicitly by (\ref{e.alpha})) such
that:
\[
\begin{array}{l}
\jap{\psi,[\tH,iA_{\alpha}]\psi}>0,
\\[3mm]
\jap{\psi,[\lambda r^{-p+1}|\psi|^{p-1},iA_{\alpha}]\psi}>0
\end{array}
\]
for all $\psi\in \calth$, where
$A_\alpha=\half((\rs-\alpha)\Drs+\Drs(\rs- \alpha))$.
\label{dil.thm1}
\end{proposition}

\proof
We have:
\[
i[\tH,A_\alpha]=2D_\rs^2-(\rs-\alpha)\frac{dV(\rs)}{d\rs},
\]
where
\[
\frac{dV(\rs)}{d\rs}=\frac{2M}{r^4}\bl 1-\frac{2M}{r}\br
\bl \frac{8M}{r}-3\br
\]
is positive for $2M<r<8M/3$ and negative for $r>8M/3$. Let
\beq
\alpha=\frac{8M}{3}+2M\log\frac{2M}{3}
\label{e.alpha}
\eeq
be the value of $\rs$ corresponding to $r=8M/3$. Then $-(\rs-\alpha)
\frac{dV}{d\rs} > 0$ for all $\rs\in\rr$, $\rs\neq\alpha$; hence:
\[
\int_{-\infty}^\infty \bar\psi\cdot[\tH,iA_\alpha]\psi\,d\rs>0
\]
for all $\psi\in\calth$, $\psi\not\equiv 0$.

Next, we consider the commutator with the nonlinear term:
\beq
\begin{array}{l}
\intt\bar\psi\cdot[\lambda r^{-p+1}|\psi|^{p-1},iA_\alpha]\psi\,d\rs
\\[3mm]
=-\lambda\intt
|\psi|^2(\rs-\alpha)\drs\big(r^{-p+1}|\psi|^{p-1}\big)d\rs.
\end{array}
\label{dil.e1}
\eeq
We have
\beq
|\psi|^2\drs(r^{-p+1}|\psi|^{p-1})
=\frac{p-1}{p+1}\,r^2\drs(r^{-p-1}|\psi|^{p+1}).
\label{dil.e2}
\eeq
Using (\ref{dil.e2}) and integrating the right-hand side of 
(\ref{dil.e1}) by parts, we obtain that it is equal to:
\beq
\frac{\lambda(p-1)}{p+1}\intt\drs(r^2(\rs-\alpha))\cdot
\big(r^{-p-1}|\psi|^{p+1}\big)d\rs.
\label{dil.e3}
\eeq
Since $r$ is a positive and increasing function of $\rs$, so is
$r^2(\rs-\alpha)$, hence $\p_{\rs}(r^2(\rs-\alpha))\geq 0$ and the
integrand in (\ref{dil.e3}) is nonnegative. Since $\lambda>0$,
(\ref{dil.e3}) is nonnegative. This completes the proof of the 
proposition.
\qed


\section{Local decay}
\label{dec}
\init


The purpose of this section is to prove the following local decay
estimates.

\begin{proposition}
Suppose that $\psi_t$ solves (\ref{red.e2}), $\psi_0\in\tilde\calh$,
and let $\beta>3/2$, $0\leq R<\infty$. Then
\beq
\int_{-\infty}^\infty \|(1+\rs^2)^{-\beta/2}\psi_t\|^2dt
\leq C_\beta\|\psi_0\|_{L^2(\rr;d\rs)}\|\psi_0\|_{\tilde\calh},
\label{dec.e13}
\eeq
\beq
\int_{-\infty}^\infty dt \int_{-R}^R r^{-p-1}|\psi_t|^{p+1}d\rs
\leq C_R\|\psi_0\|_{L^2(\rr;d\rs)}\|\psi_0\|_{\tilde\calh}.
\label{dec.e14}
\eeq
\label{dec.thm2}
\end{proposition}

\bigskip

Proposition \ref{dec.thm2} will be obtained as a consequence
of Proposition \ref{dec.thm1} below.

\bigskip

\begin{proposition}
Let $\tg(\rs)=g(\rs-\alpha)$, where $g(s)=\int_0^s(1+t^2)^{-\sigma}dt$
for some $\sigma\in(1/2,3/2)$ and $\alpha$ is as in Proposition 
\ref{dil.thm1}. Define:
\beq
\gamma=\half(\tg(\rs)D_\rs+D_\rs\tg(\rs)).
\label{dec.e2}
\eeq

(i) Let $\psi_t$ solve (\ref{red.e2}), $\psi_0\in\tilde\calh$.  Then
$\|\gamma\psi_t\|_{L^2(\rr;d\rs)}\leq C\|\psi_0\|_{\tilde\calh}$
uniformly in $t$.

\smallskip
(ii) For any $0<R<\infty$ there are $c_1, c_{2,R}>0$, such that
for all $\psi\in \calth$:
\beq
\jap{\psi, i[\hp,\gamma]\psi}
\geq c_1\intt(1+\rs^2)^{-\sigma-1}|\psi|^2\,d\rs
+ c_{2,R}\int_{-R}^R r^{-p-1}|\psi|^{p+1}d\rs.
\label{dec.e3}
\eeq
\label{dec.thm1}
\end{proposition}

{\it Proof of Proposition \ref{dec.thm2}, given Proposition 
\ref{dec.thm1}.}
Clearly, it suffices to prove the proposition for $3/2<\beta<5/2$.
By Proposition \ref{dec.thm1}{\it(ii)}, we have:
\[
\begin{array}{l}
\frac{d}{dt}\jap{\psi_t,\gamma\psi_t}
=\jap{\psi_t,i[\hp,\gamma]\psi_t}
\\[3mm]
\geq c_1\jap{\psi_t,(1+\rs^2)^{-\beta}\psi_t}
+c_{2,R} \int_{-R}^R r^{-p-1}|\psi_t|^{p+1}d\rs,
\end{array}
\]
where $\beta=\sigma+1$. Integrating this inequality from $-\infty$
to $\infty$ in $t$, we obtain:
\[
\begin{array}{l}
\int_{-\infty}^\infty\jap{\psi_t,(1+\rs^2)^{-\beta}\psi_t}
+\int_{-\infty}^\infty dt \int_{-R}^R r^{-p-1}|\psi_t|^{p+1}d\rs
\\[3mm]
\leq\overline{\lim}_{t\to \infty}\jap{\psi_t,\gamma\psi_t}
-\underline{\lim}_{t\to-\infty}\jap{\psi_t,\gamma\psi_t}
\\[3mm]
\leq 2\sup_{t\in\rr}|\jap{\psi_t,\gamma\psi_t}|
\\[3mm]
\leq 2\|\psi_t\|_{L^2}\,\|\psi_t\|_{\tilde\calh}
\leq 2\|\psi_0\|_{L^2}\,\|\psi_0\|_{\tilde\calh},
\end{array}
\]
by Proposition \ref{dec.thm1}{\it(i)}.
\qed

\bigskip

{\it Proof of Proposition \ref{dec.thm1}.}
We first note that $\tg$ is bounded if $\sigma>1/2$.  Hence:
\[
\|\gamma\psi_t\|_{L^2(\rr;d\rs)}\leq C\|\psi_t\|_{H^1(\rr;d\rs)},
\]
which implies {\it(i)}.

To prove {\it(ii)}, it suffices to verify that:
\beq
\jap{\psi, i[\tH,\gamma]\psi}
\geq c_1\intt(1+\rs^2)^{-\sigma-1}|\psi|^2\,d\rs,
\label{dec.e3a}
\eeq
\beq
\jap{\psi, i[\lambda r^{-p+1}|\psi|^{p-1},\gamma]\psi}
\geq c_{2,R}\int_{-R}^R r^{-p-1}|\psi|^{p+1}d\rs.
\label{dec.e3b}
\eeq
The proof of (\ref{dec.e3b}) is similar to that of
Proposition \ref{dil.thm1}. We have:
\[
i[\Psi(\rs),\gamma]=-\tg \drs\Psi(\rs).
\]
Putting $\Psi(\rs)=\lambda r^{-p+1}|\psi|^{p-1}$, we obtain:
\[
\begin{array}{l}
\intt\bar\psi\cdot[\lambda r^{-p+1}|\psi|^{p-1},i\gamma]\psi\,d\rs
\\[3mm]
=-\intt |\psi|^2\tg(\rs)\drs\big(\lambda r^{-p+1}|\psi|^{p-1}\big)d\rs
\\[3mm]
=-\intt\tg(\rs)\frac{\lambda(p-1)}{p+1}r^2
\drs\big(r^{-p-1}|\psi|^{p+1}\big)d\rs
\\[3mm]
=\frac{\lambda(p-1)}{p+1}\intt\drs(r^2\tg(\rs))\cdot
\big(r^{-p-1}|\psi|^{p+1}\big)d\rs,
\end{array}
\]
where we used (\ref{dil.e2}) and, at the last step, integrated by
parts. We now use that for any $R>0$ there is an $\epsilon>0$
such that
\[
\drs\bigg(r^2\tg(\rs)\bigg)\geq \epsilon
\]
for $-R\leq\rs\leq R$, and conclude that the last integral is
\[
\geq \epsilon\int_{-R}^R r^{-p-1}|\psi|^{p+1}d\rs.
\]

The proof of (\ref{dec.e3a}) essentially follows \cite{L}. We compute:
\beq
i[D_\rs^2,\gamma]=2D_\rs\tg D_\rs-\half\tg''',
\label{dec.e4}
\eeq
\beq
i[V,\gamma]=-\tg \frac{d}{d\rs}V(\rs).
\label{dec.e5}
\eeq
Since $\tg(\rs)>0$ for $\rs>\alpha$ and $\tg(\rs)<0$ for $\rs<\alpha$,
and the opposite inequalities hold for $V'(\rs)$ (see Section
\ref{dil}), the term (\ref{dec.e5}) is nonnegative. 
It remains to prove that:
\beq
\int\bar\psi\cdot i[D_{\rs}^2,\gamma]\psi d\rs
\geq\int(1+\rs^2)^{-\sigma-1}|\psi|^2d\rs.
\label{dec.e6}
\eeq
We first define the unitary transformation:
\[
\begin{array}{c}
S:\ L^2(\rr;d\rs)\to L^2(\rr;s^2ds),
\\[3mm]
\psi(\rs)\to s^{-1}\psi(s+\alpha)=:\phi(s).
\end{array}
\]
Then:
\beq
\begin{array}{l}
Si[D_\rs^2,\gamma]S^*=-2s^{-1}\dds\, g'(s)\dds\, s-\half g'''(s)
\\[3mm]
=-2\dds\, g' \dds-\frac{4}{s}g'\dds-\frac{2}{s}g''-\half g'''.
\end{array}
\label{dec.e7}
\eeq
Since
\[
\int\bar\psi\cdot i[D_{\rs}^2,\gamma]\psi\,d\rs
=\int\bar\phi\cdot S i[D_{\rs}^2,\gamma]S^*\phi\,s^2 ds,
\]
(\ref{dec.e6}) is equivalent to
\beq
\int\bar\phi\cdot Si[D_{\rs}^2,\gamma]S^*\phi\,s^2 ds
\geq c_1\int(1+\rs^2)^{-\sigma-1}|\phi|^2s^2ds.
\label{dec.e8}
\eeq
It therefore suffices to prove (\ref{dec.e8}) for $\phi\in L^2(\rr;
s^2ds)$.

We first prove that the operator
\beq
L=-\dds g'\dds-\frac{2}{s}g'\dds
\label{dec.e10}
\eeq
is nonnegative on $L^2(\rr; s^2ds)$. Writing $\int_{-\infty}^\infty
\bar\phi\,L\phi\,s^2ds =\int_{-\infty}^0\ +\int_0^\infty$, and
changing variables $s\to -s$ in the integral $\int_{-\infty}^0$,
we see that it suffices to prove that
\beq
\int_0^\infty \bar\phi\,L\phi\,s^2ds>0
\label{dec.e11}
\eeq
for $\phi\in\calh_1:=L^2([0,\infty); s^2ds)$. (We denote here by $L$
the operator defined in (\ref{dec.e10}) acting on $\calh_1$.)
To this end, we observe that $\calh_1$ can be identified with the
subspace $\calh_2$ of $L^2(\rr^3; d^3x)$ consisting of spherically
symmetric functions. Namely, we introduce spherical coordinates
$(s,\omega)$ in $\rr^3$ so that $s^2=x_1^2+x_2^2+x_3^2$, $d^3x=s^2
ds$. Then the operator
\[
\begin{array}{c}
T:\calh_1\to \calh_2,
\\[3mm]
\phi(s)\to \tilde\phi(s,\omega)=\phi(s),
\end{array}
\]
is unitary. Under this identification, $L$ becomes an operator
on $\calh_2$ equal to $TLT^*$. However, an explicit computation
shows that
\[
TLT^*=\sum_{i=1}^3 D_{x_i}\frac{x_i}{s}g'(s)\frac{x_i}{s}D_{x_i},
\]
which is obviously nonnegative since $g'=(1+s^2)^{-\sigma}>0$. Hence
$L$ is nonnegative.

To finish the proof of (\ref{dec.e8}), it remains to check that
\beq
-\frac{2}{s}g''-\half g'''\geq c_0(1+s^2)^{-\sigma-1}
\label{dec.e12}
\eeq
for some $c_0>0$. However, by an explicit computation the left-hand
side of (\ref{dec.e12}) is equal to
\[
\sigma(1+s^2)^{-\sigma-2}(5+(3-2\sigma)s^2),
\]
so that (\ref{dec.e12}) holds if $0<\sigma<3/2$.
\qed


\section{Pseudoconformal identity}
\label{conf}
\init


Let $\calx=\{\phi\in{\calth}:\ \|\phi\|_\calx<\infty\}$, where:
\[
\|\phi\|_\calx=\|\phi\|_{\calth}+\|\rs\phi\|_{L^2(\rr;d\rs)}.
\]
We continue to denote by $\jap{\cdot,\cdot}$ the inner product in
$L^2(\rr; d\rs)$.

\begin{proposition}
Assume that $p>3$, and let $\eps>0$. Let $\psi_t$ be a solution of
(\ref{red.e2}) such that $\psi_1 \in\calx$. Then:
\beq
\int_{1}^T\jap{\psi_t,\bl\frac{\rs}{2t}-D_\rs\br^2\psi_t}dt
<CT^\eps(\|\psi_1\|^2_\calx+\|\psi_1\|^{p+1}_\calth);
\label{conf.e1}
\eeq
\beq
\int_1^T \int_{-\infty}^\infty r^{-p+1}|\psi_t|^{p+1}d\rs dt
<CT^\eps(\|\psi_1\|^2_\calx+\|\psi_1\|^{p+1}_\calth);
\label{conf.e2}
\eeq
and, for $t>1$,
\beq
\jap{\psi_t,\bl \frac{\rs}{2t}-D_\rs\br^2\psi_t}
<Ct^{-1+\eps}(\|\psi_1\|^2_\calx+\|\psi_1\|^{p+1}_\calth);
\label{conf.e3}
\eeq
\beq
\int_{-\infty}^\infty r^{-p+1}|\psi_t|^{p+1}d\rs 
<Ct^{-1+\eps}(\|\psi_1\|^2_\calx+\|\psi_1\|^{p+1}_\calth);
\label{conf.e4}
\eeq
\beq
\jap{\psi_t,V(\rs)\psi_t}
<Ct^{-1+\eps}(\|\psi_1\|^2_\calx+\|\psi_1\|^{p+1}_\calth);
\label{conf.e5}
\eeq
the constants in (\ref{conf.e1})--(\ref{conf.e5}) may depend on 
$\lambda$, $p$, $\eps$ but are independent of $T$, $t$.
\label{conf.thm1}
\end{proposition}

\proof 
Throughout this proof we will denote by $C$ a constant which may depend
on $p$, $\lambda$, $\eps$ and may change from line to line, but is
always independent of $\psi$, $t$, $T$.

Let
\[
\Phi(t)=\Phi_0(t)+t\Psi(t),
\]
where
\[
\Phi_0(t)=t\bl\bl\frac{\rs}{2t}-D_\rs\br^2+V\br,\ 
\Psi(t)=\lambda r^{-p+1}|\psi_t|^{p-1}.
\]
Observe that:
\beq
0\leq\jap{\psi_1,\Phi_0(1)\psi_1}\leq C\|\psi_1\|^2_\calx,
\label{co.e1}
\eeq
\beq
0\leq\jap{\psi_1,\Psi(1)\psi_1}\leq C\|\psi_1\|^{p+1}_\calth.
\label{co.e2}
\eeq
Indeed, (\ref{co.e1}) is obvious from the definition of $\Phi_0$ and
$\calx$. To prove (\ref{co.e2}), we will use that in dimension $1$:
\beq
\|\psi\|_\infty\leq C\|\psi\|_2^{1/2}\|D_\rs\psi\|_2^{1/2},
\label{onedim.e}
\eeq
and that $r^{-1}$ is bounded, hence:
\[
\jap{\psi_1,\Psi_1\psi_1}
\leq C\int_{-\infty}^\infty |\psi_1|^{p+1}d\rs
\leq C\|\psi_1\|^{p-1}_\infty\int_{-\infty}^\infty |\psi_1|^2d\rs
\leq C\|\psi_1\|^{p+1}_\calth.
\]

The main idea of the proof of Proposition \ref{conf.thm1} is to estimate
\[
\jap{\psi_T,\Phi(T)\psi_T}-\jap{\psi_0,\Phi(0)\psi_0}
=\int_1^T \frac{d}{dt}\jap{\psi_t,\Phi(t)\psi_t}dt
\]
from below. We have $
\frac{d}{dt}\jap{\psi_t,\Phi(t)\psi_t}=\jap{\psi_t,D\Phi(t)\psi_t},
$ where
\[
D\Phi(t)=\frac{\p}{\p t}\Phi(t)+i[\tH_{\psi},\Phi(t)].
\]
We will also write
\[
D_0\Phi_0(t)=\frac{\p}{\p t}\Phi_0(t)+i[\tH,\Phi_0(t)]
\]
We compute:
\beq
\begin{array}{l}
D\Phi(t)=\frac{\p}{\p t}(\Phi_0(t)+t\Psi)+i[\tH+\Psi,\Phi_0(t)+t\Psi]
\\[3mm]
=D_0\Phi_0(t)+\frac{\p}{\p t}(t\Psi)+i[\tH,t\Psi]+i[\Psi,\Phi_0]
\\[3mm]
=D_0\Phi_0(t)+\frac{\p}{\p t}(t\Psi)+i[D_\rs^2,t\Psi]
+i[\Psi,t\bl\frac{\rs}{2t}-D_\rs\br^2]
\\[3mm]
=D_0\Phi_0(t)+\frac{\p}{\p t}(t\Psi)+i[\Psi,-\half(\rs D_\rs+D_\rs\rs)]
\\[3mm]
=D_0\Phi_0(t)+\Psi+t\frac{\p}{\p t}\Psi+\rs \frac{\p}{\p\rs}\Psi(\rs).
\end{array}
\label{conf.e6}
\eeq

We first estimate the nonlinear terms, beginning with $\jap{\psi_t,\rs
\frac{\p}{\p\rs}\Psi(\rs)\psi_t}$:
\beq
\begin{array}{l}
\intt|\psi_t|^2\rs\drs\Psi d\rs
=\intt\frac{\lambda(p-1)}{p+1}r^2\rs\drs(r^{-p-1}|\psi_t|^{p+1})d\rs
\\[3mm]
=-\frac{\lambda(p-1)}{p+1}\intt\drs(r^2\rs)\,r^{-p-1}|\psi_t|^{p+1}
d\rs
\\[3mm]
=-\frac{\lambda(p-1)}{p+1}\intt(\drs(r^2)\rs+r^2)r^{-p-1}|\psi_t|^{p+1}
d\rs.
\\[3mm]
=-\frac{\lambda(p-1)}{p+1}\Big(\int_{-\infty}^{-R}+\int_{-R}^R
+\int_R^\infty\Big)\drs(r^2)\rs r^{-p-1}|\psi_t|^{p+1} d\rs
\\[3mm]
-\frac{\lambda(p-1)}{p+1}\intt r^{-p+1}|\psi_t|^{p+1} d\rs.
\end{array}
\label{conf.e8}
\eeq
By (\ref{ham.e2}), for any $\delta>0$ there is $R>0$ such that
$|\drs(r^2)\rs r^{-2}|<\lambda\delta$ for $\rs<-R$. Hence
\beq
\int_{-\infty}^{-R}|\drs(r^2)\rs| r^{-p-1}|\psi_t|^{p+1} d\rs
\leq\lambda\delta\intt r^{-p+1}|\psi_t|^{p+1} d\rs,
\label{conf.e9}
\eeq
provided that $R$ is large enough. Next, $\drs(r^2)\rs$ is bounded
for $-R\leq\rs\leq R$, so that:
\beq
\begin{array}{l}
\int_{-\infty}^\infty dt 
\int_{-R}^{R}|\drs(r^2)\rs| r^{-p-1}|\psi_t|^{p+1} d\rs
\\[3mm]
\leq C\int_{-\infty}^\infty dt\intt r^{-p-1}|\psi_t|^{p+1} d\rs
\leq C\|\psi_1\|^2_\calx,
\end{array}
\label{conf.e10}
\eeq
by (\ref{dec.e13}); the constant $C$ depends on
$R$ and hence on $\delta$, but not on $T$. Finally,
\beq
\int_R^\infty \drs(r^2)\rs r^{-p-1}|\psi_t|^{p+1} d\rs\geq 0.
\label{conf.e11}
\eeq
Plugging (\ref{conf.e9})--(\ref{conf.e11}) into (\ref{conf.e8}),
we obtain:
\beq
\begin{array}{l}
\int_{1}^T \jap{\psi_t,\rs\drs\Psi(t)\psi_t}dt
\\[3mm]
\leq -\lambda\Big(\frac{p-1}{p+1}-\delta\Big)
\int_1^T dt \int r^{-p+1}|\psi_t|^{p+1} d\rs
+ C\|\psi_1\|^2_\calx
\\[3mm]
= -(\frac{p-1}{p+1}-\delta)
\int_1^T \jap{\psi_t,\Psi(t)\psi_t}+C\|\psi_1\|^2_\calx.
\end{array}
\label{conf.e12}
\eeq
Next, we have
\beq
\begin{array}{l}
\int_1^T t\jap{\psi_t,\frac{\partial \Psi}{\partial t}\psi_t}dt
=\lambda \int_1^T t\intt|\psi_t|^2\frac{\partial}{\partial t}
(r^{-p+1}|\psi_t|^{p-1})d\rs dt
\\[3mm]
=\frac{\lambda(p-1)}{p+1}\int_1^T t\intt\frac{\partial}{\partial t}
(r^{-p+1}|\psi_t|^{p+1})d\rs dt
\\[3mm]
=\frac{p-1}{p+1}\int_1^T t\frac{\partial}{\partial t}
\jap{\psi_t,\Psi(t)\psi_t} dt
\\[3mm]
=\frac{p-1}{p+1}\Big(T\jap{\psi_T,\Psi(T)\psi_T}
-\jap{\psi_1,\Psi(1)\psi_1}
-\int_1^T \jap{\psi_t,\Psi(t)\psi_t}\Big) dt.
\end{array}
\label{conf.e13}
\eeq
Combining (\ref{conf.e12}) and (\ref{conf.e13}), we obtain:
\beq
\begin{array}{l}
\int_1^T\jap{\psi_t,\Big(\Psi(t)+t\frac{\p}{\p t}\Psi(t)
+\rs\drs\Psi(t)\Big)\psi_t}\,dt
\\[3mm]
\leq \frac{p-1}{p+1}\Big(T\jap{\psi_T,\Psi(T)\psi_T}
-\jap{\psi_1,\Psi(1)\psi_1}\Big)
-c_1\int_1^T \jap{\psi_t,\Psi(t)\psi_t}dt
+C\|\psi_1\|^2_\calx,
\end{array}
\label{conf.e14}
\eeq
where $c_1=\frac{p-3}{p+1}-\delta$ satisfies $c_1>0$ if $\delta>0$ was
chosen small enough.

It remains to estimate:
\[
D_0\Phi_0=-\Big(\frac{\rs}{2t}-D_{\rs}\Big)^2+W(\rs),
\]
where
\[
W(\rs)=V(\rs)+\rs\drs V(\rs)
=\frac{2M}{r^3}\Big(1-\frac{2M}{r}\Big)
+\frac{2M\rs}{r^4}\Big(1-\frac{2M}{r}\Big)\Big(\frac{8M}{r}-3\Big).
\]
We have:
\[
|W(\rs)|\leq C(1+r^2_*)^{-3/2},
\]
which is almost -- but not quite -- sufficient for the local decay
estimate (\ref{dec.e13}) to be applicable.  To remedy this,
we write:
\[
\int_1^T\jap{\psi_t,W(\rs)\psi_t}dt
\leq C\int_1^T\jap{\psi_t,(1+r^2_*)^{-3/2}\,\psi_t}dt
\leq I_1+I_2,
\]
where:
\[
\begin{array}{l}
I_1=C\int_1^T\jap{\psi_t,\chi(1+r^2_*)^{-3/2}\,\psi_t}dt,
\\[3mm]
I_2=C\int_1^T\jap{\psi_t,(1-\chi)(1+r^2_*)^{-3/2}\,\psi_t}dt,
\end{array}
\]
and $\chi(\rs,t)$ is a bounded $C^\infty$ function such that
$\chi\equiv 1$ for $1+r^2_*\leq t$, $t\geq 1$, and
$\chi\equiv 0$ for $1+r^2_*\geq 2t$, $t\geq 1$.  To estimate
$I_2$, we use that $(1+r^2_*)^{-3/2}\leq t^{-3/2}$ on $\supp\chi$,
hence:
\[
I_2\leq C\int_1^T t^{-3/2}\|\psi_t\|^2_2\, dt\leq C\|\psi_1\|^2_\calx,
\]
where we also used that the $L^2$ norm of $\psi_t$ is constant.
It remains to estimate $I_1$. We have for any $\eps>0$:
\[
\chi(1+r^2_*)^{-\frac{3}{2}}
=\chi(1+r^2_*)^{\eps}\,(1+r^2_*)^{-\frac{3}{2}-\eps}
\leq Ct^\eps(1+r^2_*)^{-\frac{3}{2}-\eps}
\leq CT^\eps(1+r^2_*)^{-\frac{3}{2}-\eps}
\]
for $1\leq t\leq T$. Hence using (\ref{dec.e13}) we obtain:
\[
I_2\leq CT^\eps\int_1^T\jap{\psi_t,(1+r^2_*)^{-3/2-\eps}\psi_t}dt
\leq CT^\eps\|\psi_1\|^2_\calx.
\]
Thus:
\beq
\int_1^T\jap{\psi_t,D_0\Phi_0\psi_t}dt
\leq -\int_1^T\jap{\psi_t,\Big(\frac{\rs}{2t}-D_{\rs}\Big)^2\psi_t}dt
+ CT^\eps\|\psi_1\|^2_\calx.
\label{co.e3}
\eeq

Combining (\ref{co.e3}) and (\ref{conf.e14}), we find that:
\beq
\begin{array}{l}
\jap{\psi_T,\big(\Phi_0(T)+T\Psi(T)\big)\psi_T}
-\jap{\psi_1,\big(\Phi_0(1)+\Psi(1)\big)\psi_1}
\\[3mm]
=\int_1^T\jap{\psi_t,D\Phi(t)\psi_t}dt
\\[3mm]
\leq -\int_1^T\jap{\psi_t,\Big(\frac{\rs}{2t}-D_{\rs}\Big)^2\psi_t}dt
-c_1\int_1^T\jap{\psi_t,\Psi(t)\psi_t}dt
\\[3mm]
+\frac{p-1}{p+1}\Big(T\jap{\psi_T,\Psi(T)\psi_T}
-\jap{\psi_1,\Psi(1)\psi_1}\Big)
+ CT^\eps\|\psi_1\|^2_\calx.
\end{array}
\label{co.e4}
\eeq
Rearranging (\ref{co.e4}) and using (\ref{co.e1})--(\ref{co.e2}),
we finally obtain:
\beq
\begin{array}{l}
\jap{\psi_T,\Phi_0(T)\psi_T}+\frac{2}{p+1}\,T\jap{\psi_T,\Psi(T)\psi_T}
\\[3mm]
\leq -\int_1^T\jap{\psi_t,\Big(\frac{\rs}{2t}-D_{\rs}\Big)^2\psi_t}dt
-c_1\int_1^T\int r^{-p+1}|\psi_t|^{p+1}\,d\rs dt
\\[3mm]
+ C\Big(T^\eps\|\psi_1\|^2_\calx+\|\psi_1\|^{p+1}_\calth\Big).
\end{array}
\label{co.e5}
\eeq

Note first that the integrands in the integrals
\beq
\begin{array}{l}
\int_{1}^T\jap{\psi_t,\bl\frac{\rs}{2t}-D_\rs\br^2\psi_t}dt,
\\[3mm]
\int_1^T\int r^{-p+1}|\psi_t|^{p+1}d\rs dt
\end{array}
\label{conf.e16}
\eeq
are positive. Hence the left-hand side of (\ref{co.e5}) is bounded
from above, uniformly in $T$, by
\[
C\Big(T^\eps\|\psi_1\|^2_\calx+\|\psi_1\|^{p+1}_\calth\Big).
\]
However, it is also trivially bounded from below by $0$, since
$\Phi_0(T)$ and $\Psi(T)$ are nonnegative for all $T$. 
This yields (\ref{conf.e3})--(\ref{conf.e5}).
Finally, to get (\ref{conf.e1})--(\ref{conf.e2}) we plug
(\ref{conf.e3})--(\ref{conf.e5}) back into (\ref{co.e5}). 
This ends the proof of the proposition.
\qed


\section{Global existence and scattering theory}
\label{es}
\init


In this section we prove our main $L^\infty$ estimate on $\psi_t$
(Proposition \ref{es.prop1}). Interpolating this estimate with
the conservation of the $L^2$ norm yields $L^p$ estimates, which
will be an essential part of our proof of asymptotic completeness.
Recall that the space $\calx$ was defined at the beginning of
Section \ref{conf}.

\begin{proposition}
Let $\psi_t$ solve (\ref{red.e2}), $\psi_0\in\calx$.
Assume that $\lambda>0$ and $p>3$. Then for any $\eps>0$:
\beq
\|\psi_t\|_\infty\leq C_\eps\,t^{-\frac{1}{4}+\eps} \|\psi_1\|_2^{1/2}
\Big(\|\psi\|_\calx^2+\|\psi\|_\calth^{p+1}\Big)^{1/4}.
\label{es.e1}
\eeq
(This in particular implies global existence in $L^\infty$.)
\label{es.prop1}
\end{proposition}

\proof
>From (\ref{onedim.e}) we have:
\[
\begin{array}{l}
\|\psi_t\|_\infty=\|e^{ix^2/4t}\psi_t\|_\infty
\leq C\|\psi_t\|_2^{1/2} \|D_\rs e^{ix^2/4t}\psi_t\|_2^{1/2}
\\[3mm]
= C\|\psi_t\|_2^{1/2} \|(\frac{x}{2t}-D_\rs)\psi_t\|_2^{1/2}
\\[3mm]
\leq Ct^{-\frac{1}{4}+\eps}\|\psi_0\|_2^{1/2}
\Big(\|\psi\|_\calx^2+\|\psi\|_\calth^{p+1}\Big)^{1/4}.
\end{array}
\]
where we have used that, by (\ref{conf.e3}),
\[
\|(\frac{x}{2t}-D_\rs)\psi_t\|_2\leq Ct^{-\frac{1}{2}+\eps}
\Big(\|\psi\|_\calx^2+\|\psi\|_\calth^{p+1}\Big)^{1/2}.
\]
This proves (\ref{es.e1}).
\qed

\bigskip

The main result of this section is the following theorem, in which we
compare the nonlinear dynamics associated with the equation
(\ref{red.e2}) to the linear evolution $e^{-it\tH}$. We will
state all of our results on scattering for the case $t\to\infty$;
for $t\to -\infty$ analogous results obviously hold.

\begin{theorem}
(i) (Existence of wave operators)  
Assume that $p>\half(3+\sqrt{17})(\approx 3.56)$. Then for any
$\psi_+\in H^1(\rr)\cap L^{q'}(\rr)$, where $q'=1+\frac{1}{p}$,
there is a $\psi_0\in H^1(\rr)$ such that if $\psi_t$ is the
solution of (\ref{red.e2}) with initial condition $\psi_0$ at $t=0$,
then:
\beq
\|e^{-it\tH}\psi_+-\psi_t\|_{L^2(\rr)}\to 0 \hbox{ as }t\to\infty.
\label{es.e2}
\eeq
(ii) (Asymptotic completeness)
Assume that $p>4$, and let $\psi_0\in \calx$.
Let $\psi_t$ be the solution of (\ref{red.e2}) with initial condition
$\psi_0$ at $t=0$. Then there is a $\psi_+\in L^2(\rr)$ such that
(\ref{es.e2}) is satisfied.

\label{es.thm2}
\end{theorem}

Observe that by H\"older's inequality:
\[
\int|\psi|^{q'}d\rs
\leq \bl\int|\psi|^2(\rs^2+1)\,d\rs\br^{q'/2}
\bl\int(\rs^2+1)^{-\frac{q'}{2-q'}}d\rs\br^{1-\frac{q'}{2}};
\]
the last integral is convergent since $\frac{2q'}{2-q'}>1$.
Hence $\calx\subset L^{q'}(\rr)\cap H^1(\rr)$, and 
{\it(i)} holds in particular for $\psi_+\in \calx$.

\bigskip
{\it Proof of Theorem \ref{es.thm2}.}
To prove the existence of wave operators (part {\it(i)}), we need to
solve the integral equation
\beq
\psi_t=e^{-it\tH}\psi_+-i\lambda\int_t^\infty e^{-i(t-s)\tH}
r^{-p+1}|\psi_s|^{p-1}\psi_s\,ds
\label{es.e3}
\eeq
for a given $\psi_+\in H^1(\rr)\cap L^2(\rr;\rs d\rs)$. Let
\beq
\calf(\phi)=\int_t^\infty e^{-i(t-s)\tH}r^{-p+1}|\phi_s|^{p-1}\phi_s\,ds.
\label{es.e4}
\eeq
Let $X_T=L^k([T,\infty); L^q(\rr))$ for suitable $k,q$ which
will be chosen later. We first prove that
\beq
\|\calf(\phi)\|_{X_T}\leq C_0\|\phi\|^p_{X_T},
\label{es.e5}
\eeq
with $C_0$ independent of $\phi$ and $T$.

We will use the $L^q$ estimates for the Schr\"odinger unitary 
group $e^{-itH}$ in dimension 1, proved recently by Weder
\cite{W}: if $H=D^2+V(x)$ is a Schr\"odinger
operator on $L^2(\rr)$ with the potential $V(x)$ satisfying
$\int|V(x)|(1+|x|)^\gamma dx<\infty$ for some $\gamma>5/2$
(which clearly holds for $V$ given by (\ref{red.e1}), then:
\beq
\|e^{-itH}P_c\|_{B(L^{q'},L^q)} \leq Ct^{-(\half-\frac{1}{q})}
\label{es.e6}
\eeq
for $1\leq q'\leq 2$, $\frac{1}{q}+\frac{1}{q'}=1$.  $P_c$ is the
projection on the continuous spectral subspace of $H$; for $H=\tH$,
it follows \eg from Proposition \ref{dil.thm1} that $\tH$ has no point
spectrum and therefore $P_c={\bf 1}$.

Using (\ref{es.e6}) and the fact that $r^{-1}$ is bounded, we estimate:
\beq
\begin{array}{l}
\|\calf(\phi)\|_{X_T}
=\Big\|\,\|\int_t^\infty e^{-i(t-s)H}r^{-p+1}|\phi_s|^{p-1}\psi_s\,ds
\|_{L^q(d\rs)}\Big\|_{L^k(dt)}
\\[3mm]
\leq\Big\|\int_t^\infty|t-s|^{-(\half-\frac{1}{q})}
\|r^{-p+1}\phi_s^{p}\|_{L^{q'}(d\rs)}ds\Big\|_{L^k(dt)}
\\[3mm]
\leq\Big\|\int_t^\infty|t-s|^{-(\half-\frac{1}{q})}
\|\phi_s\|^p_{L^{pq'}(d\rs)}ds\Big\|_{L^k(dt)}
\\[3mm]
\leq C_0\Big\|\,\|\phi_s\|_{L^{pq'}(d\rs)}\Big\|^p_{L^{p\eta}(dt)},
\end{array}
\label{es.e7}
\eeq
where:
\beq
1+\frac{1}{k}=\frac{1}{\kappa}+\frac{1}{\eta},\ 
\frac{1}{q}+\frac{1}{q'}=1,\ 
\Big(\half-\frac{1}{q}\Big)\kappa=1
\label{es.e8}
\eeq
(at the last step we used the generalized Young's inequality). The
double norm in
the last line of (\ref{es.e7}) is equal to $\|\phi\|_{X_T}^p$ if
\beq
p\eta=k,\ pq'=q.
\label{es.e9}
\eeq
Solving (\ref{es.e8})--(\ref{es.e9}), we obtain:
\beq
q=p+1,\ \kappa=\frac{2(p+1)}{p-1},\ 
k=\frac{2(p-1)(p+1)}{p+3}. 
\label{es.e10}
\eeq
Hence, if $q,k$ are chosen as in (\ref{es.e10}), the
mapping $\calf$ is a contraction on the ball
$B_{\epsilon,T}=\{\|\phi\|_{X_T}\leq\epsilon\}$, where $\epsilon$
depends on $p$ but not on $T$.

Using (\ref{es.e6}), we obtain that for $\psi_+\in L^2(\rr)\cap
L^{q'}(\rr)$,
\[
\big\|\|e^{-it\tH}\psi_+\|_{L^q(d\rs)}\big\|_{L^k([T,\infty);dt)}
\to 0 \hbox{ as } T\to\infty,
\]
provided that $(\half-\frac{1}{q})k>1$. For $q$ and $k$ as in
(\ref{es.e10}), this condition is satisfied when:
\[
p>\frac{3+\sqrt{17}}{2}\approx 3.56.
\]
Therefore, given a $\psi_+\in H^1(\rr)\cap L^{q'}(\rr)$, we may choose 
$T$ large enough so that $e^{-it\tH}\psi_+\in B_{\epsilon/3, T}$.
Using a standard contraction argument, we can now solve (by
iteration) the equation (\ref{es.e3}) for $t\geq T$.
The solution $\psi_t(\rs), t\geq T,$ belongs to $X_{T}$ and
solves (in the weak sense) the 
differential equation (\ref{red.e2}) with the initial condition
$\psi_{T}=e^{-iT\tH}\psi_+\in H^1(\rr)$. By conservation of
energy, $\psi_t\in H^1(\rr)$ for $t\geq T$. Finally, we extend
$\psi_t$ to all $t\in\rr$ by solving (\ref{red.e2}) backwards
(\ie for $t<T$) with the initial data as above at $t=T$.
We obtain a solution $\psi_t$ with 
\[
\|\psi_t\|_{H^1(\rr)}
\leq C\|e^{-it_0\tH}\psi_+\|_{H^1(\rr)}
\leq C\|\psi_+\|_{H^1(\rr)}
\]
for all $t\in\rr$.

Next, we claim that
\beq
\|e^{it\tH}\psi_t-\psi_+\|_{L^q(d\rs)}\to 0\hbox{ as }t\to\infty.
\label{es.e11}
\eeq
Indeed, multiplying both sides of (\ref{es.e3}) by $e^{it\tH}$
and then proceeding as in the proof of (\ref{es.e5}),
we obtain for $t>T$:
\beq
\begin{array}{l}
\|e^{it\tH}\psi_t-\psi_+\|_{L^q(d\rs)}
=\lambda\big\|\int_t^\infty e^{-is\tH}r^{-p+1}|\psi_s|^{p-1}\psi_sds
\big\|_{L^q(d\rs)}
\\[3mm]
\leq\lambda\int_t^\infty s^{-(\half-\frac{1}{q})}
\|\psi_s\|^{p}_{L^{pq'}(d\rs)}
\\[3mm]
\leq\lambda\Big(\int_t^\infty s^{-(\half-\frac{1}{q})\xi}ds
\Big)^{1/\xi}\Big(\int_t^\infty 
\|\psi_s\|^{p\eta}_{L^{pq'}(d\rs)}ds\Big)^{1/\eta},
\end{array}
\label{es.e12}
\eeq
for $\frac{1}{\eta}+\frac{1}{\xi}=1$. The last line in (\ref{es.e12})
is bounded by $C_1(t)\|\psi\|_{X_{t_0}}^p$,
with $C_1(t)\to 0$ as $t\to\infty$, if
\[
(\half-\frac{1}{q})\xi>1,\ p\eta=k,\ pq'=q.
\]
Again, it is easy to verify that one can choose $\xi,\eta$ so
that these conditions are satisfied.

By (\ref{es.e11}), $e^{it\tH}\psi_t$ converges to $\psi_+$ strongly 
in $L^q(\rr)$; moreover, we saw earlier that the $H^1$ norms
of $e^{it\tH}\psi_t$ are bounded uniformly for all $t$. Hence
$e^{it\tH}\psi_t$ has a weak limit in $L^2$. Since the $L^2$ norm
of $e^{it\tH}\psi_t$ is constant in $t$, the $L^2$ convergence
is strong. The $L^q$ limit, $\psi_+$, belongs to $L^2$, hence
the $L^2$ limit must also be equal to $\psi_+$.

We obtain that:
\[
\|\psi_t-e^{-it\tH}\psi_+\|_{L^2(d\rs)}
=\|e^{it\tH}\psi_t-\psi_+\|_{L^2(d\rs)}\to 0,
\]
which proves Theorem \ref{es.thm2}{\it(i)}.

To prove {\it(ii)}, \ie the completeness of wave operators, we 
consider the integral equation 
\beq
e^{it\tH}\psi_t=\psi_0 -i\lambda\int_0^t e^{is\tH}
r^{-p+1}|\psi_s|^{p-1}\psi_s\,ds,
\label{es.e13}
\eeq
for $\psi_0\in \calx$. The equation (\ref{es.e13}) is equivalent to
(\ref{red.e2}) with initial condition $\psi_0$ at $t=0$.

We will prove that the integral 
\beq
\int_0^\infty e^{is\tH} r^{-p+1}|\psi_s|^{p-1}\psi_s\,ds
\label{es.e14}
\eeq
is norm convergent in $L^q(\rr)$ for some $2<q<\infty$ (depending on 
$p$). Indeed, by (\ref{es.e6}) we have:
\[
\int_0^t \|e^{is\tH} r^{-p+1}|\psi_s|^{p-1}\psi_s\|_{L^2(\rr)}ds
\leq\int_0^t s^{-(\half-\frac{1}{q})}\|\psi_s\|_{L^{q'p}(\rr)}^p ds,
\]
for $\frac{1}{q}+\frac{1}{q'}=1$, $q>2$. The last integral can be
broken up into $\int_0^1+\int_1^\infty$.
Since $p>4$ and $q'\geq 1$, $q'p>4$, so that by Sobolev's inequality:
\[
\|\psi_s\|_{L^{q'p}}<C\|\psi_s\|_{H^1(\rr)}=C\|\psi_0\|_{H^1(\rr)}.
\]
Therefore the integral $\int_0^1$ is bounded by:
\[
C\|\psi_0\|_{H^1(\rr)}\int_0^t s^{-(\half-\frac{1}{q})}ds
\leq C'\|\psi_0\|_{H^1(\rr)},
\]
since $\half-\frac{1}{q}<1$.

To prove that $\int_1^t$ is finite, it suffices to verify that
\beq
\|\psi_s\|_{L^{q'p}(\rr)}\leq C(\psi_0)\,s^{-\alpha}
\label{es.e15}
\eeq
for some $\alpha$ satisfying $\half-\frac{1}{q}+\alpha p>1$, \ie
\beq
\frac{1}{q'}+\alpha p>\frac{3}{2}.
\label{es.e16}
\eeq
We obtain (\ref{es.e15}) by
interpolating between (\ref{es.e1}) and the conservation
of the $L^2$ norm: $\|\psi_s\|_{L^2}=\|\psi_0\|_{L^2}$.
Such interpolation yields (\ref{es.e15}) for
\beq
\alpha<\frac{1}{4}(1-\theta)=\frac{q'p-2}{4q'p}.
\label{es.e100}
\eeq
We may find an $\alpha$ satisfying both (\ref{es.e100}) and (\ref{es.e16})
if $q'p+2>6q'$. It therefore suffices to take $1<q'<\frac{2}{6-p}$
if $4<p<5$ and $1<q'<2$ if $p\geq 5$.

Let $\phi_+\in L^q(\rr)$ be the $L^q$ limit of (\ref{es.e14}).
Then (\ref{es.e13}) implies that:
\[
\|e^{it\tH}\psi_t-\psi_0+i\lambda\phi_+\|_{L^q(\rr)}\to 0
\hbox{ as }t\to\infty.
\]
The same argument as in the proof of {\it(i)} proves now that
the convergence takes place also in $L^2$. Let $\psi_+=\psi_0
+i\lambda\phi_+$, then:
\[
\lim_{t\to\infty}\|e^{it\tH}\psi_t-\psi_+\|_{L^2(\rr)}
=\lim_{t\to\infty}\|\psi_t-e^{-it\tH}\psi_+\|_{L^2(\rr)}=0.
\]
This ends the proof of the theorem.
\qed

\bigskip
Recall that $\tH=D_\rs^2+V(\rs)$, where $V(\rs)$ is a short-range
$C^\infty$ potential given by (\ref{red.e1}). We can therefore apply
the well known results on short range scattering
(see \eg \cite{CFKS}) to the evolution $e^{-it\tH}$. As noted
before, $\tH$ has no point spectrum, hence the wave operators
\[
W_+=s-\lim_{t\to\infty}e^{it\tH}e^{-itD_\rs^2}
\]
exist and are complete (\ie are unitary on $L^2$).
Thus for any initial condition $\psi_+\in L^2(\rr)$ there is
a $\phi_+\in L^2(\rr)$ ($\phi_+=W_+^*\psi_+$) such that:
\[
\|e^{-it\tH}\psi_+-e^{-itD_\rs^2}\phi_+\|_{L^2(\rr)}\to 0
\hbox{ as }t\to\infty.
\]
Combining this with Theorem \ref{es.thm2}{\it(ii)}, we obtain
the following corollary.

\begin{corollary}
Assume that $p>4$, and let $\psi_t$ be the solution of
(\ref{red.e2}) with the initial condition 
$\psi_0\in H^1(\rr)\cap L^2(\rr;\rs d\rs)$ at $t=0$.
Then there is a $\phi_+\in L^2(\rr)$ such that:
\beq
\|\psi_t-e^{-itD_\rs^2}\phi_+\|_{L^2(\rr)}\to 0
\hbox{ as }t\to\infty.
\label{es.e17}
\eeq
\label{es.cor3}
\end{corollary}

Similarly, by Theorem \ref{es.thm2}{\it(i)} there is a subspace $S$
of $L^2(\rr)$ (equal to $W_+^*(H^1\cap L^{1+1/p})$) such that for any
$\phi_+\in S$ there is a $\psi_0 \in H^1(\rr)$ for which (\ref{es.e17})
holds.

Finally, let us reformulate Corollary \ref{es.cor3} in terms of
the original equation (\ref{ham.e2}), to which (\ref{red.e2})
is unitarily equivalent. Recall from Section \ref{red} that 
$\psi_t$ solves (\ref{red.e2}) if and only if
\[
u_t(r,\omega)=U\psi_t(r)=r^{-1}\psi_t(r)
\]
solves (\ref{ham.e2}). Moreover, the condition $\psi_0\in \calx$ is
equivalent to:
\beq
u\in\calh,\ \rs u\in L^2(\rr\times S^2;r^2d\rs d\omega).
\label{es.e18}
\eeq
Corollary \ref{es.cor3} states that if $u_t$ solves (\ref{ham.e2})
and the initial condition $u_0$ at $t=0$ satisfies (\ref{es.e18}),
then there is a $u_+\in L^2(\rr\times S^2;r^2d\rs d\omega)$
such that:
\beq
\|Ju_t-e^{-itD_\rs^2}Ju_+\|_{L^2(\rr;d\rs)}\to 0
\hbox{ as }t\to\infty,
\label{es.e19}
\eeq
where 
\[
J=U^{-1}:\ L^2(\rr\times S^2;r^2d\rs d\omega)\to L^2(\rr;d\rs),
\]
\[
(Ju)(r)=(U^{-1}u)(r)=ru(r,\omega)
\]
for radially symmetric $u$. (\ref{es.e19}) may be interpreted as
follows (cf. \cite{Di}, \cite{DiKa}, \cite{Ba1}). All solutions
$e^{-itD_\rs^2} Ju_+$ of the free Schr\"odinger equation on the 
cylindrical manifold $\rr\times S^2$ (with the usual metric) split up
into two parts, one of which escapes to the ``spatial infinity"
$\rs\to\infty$, the other approaches the horizon $\rs\to -\infty$.
Hence $Ju_t$, where $u_t$ solves (\ref{ham.e2}), will have similar
characteristics.  However, if we return to the usual coordinates
on the Schwarzschild manifold, the waves approaching the horizon
and the spatial infinity will begin to look differently.
As $\rs\to\infty$, $r\sim \rs$ and the asymptotic
dynamics generated by $J^{-1}D_\rs^2 J$, is similar to that
for a free Schr\"odinger equation in $\rr^3$. On the other hand,
when $\rs\to -\infty$, $r\to 2M$ and the identification operator $J$
is essentially a multiplication by $2M$; hence the asymptotic evolution
is given by a one-dimensional Schr\"odinger equation. 
This phenomenon seems to be typical for evolution equations on
Schwarzschild manifolds, cf. \cite{Ba1}, Section 1.


\end{document}